\documentstyle[12pt,aaspp4]{article}
\tightenlines

\newcommand{\be}{\begin{equation}}
\newcommand{\ee}{\end{equation}}

\newcommand{\vecE}{\mbox{\boldmath$E$}}

\newcommand{\ofxh}{\mbox{$\hat{\vecx}^{\bf\rm o}$}}
\newcommand{\ofyh}{\mbox{$\hat{\vecy}^{\bf\rm o}$}}
\newcommand{\ofzh}{\mbox{$\hat{\vecz}^{\bf\rm o}$}}

\newcommand{\vecx}{\mbox{\boldmath$x$}}      
\newcommand{\vecy}{\mbox{\boldmath$y$}}      
\newcommand{\vecz}{\mbox{\boldmath$z$}}

\newcommand{\Cpar}{\mbox{$C_{\|}$}}
\newcommand{\Cper}{\mbox{$C_{\bot}$}}
\newcommand{\Cavg}{\mbox{$C_{\rm avg}$}}

\newcommand{\Cxo}{\mbox{$C_x^{\rm o}$}}
\newcommand{\Cyo}{\mbox{$C_y^{\rm o}$}}

\def    \be     {\begin{equation}}
\def    \ee     {end{equation}}
\def    \ba     {\begin{eqnarray}}
\def    \ea     {\end{eqnarray}}
\def	\Angstrom	{\,{\rm \AA}}		

\def	\ba	{{\bf a}}

\def    \simlt  {\lower.5ex\hbox{$\; \buildrel < \over \sim \;$}}
\def    \simgt  {\lower.5ex\hbox{$\; \buildrel > \over \sim \;$}}

\def\pmb#1{\setbox0=\hbox{#1}%
\kern-.025em\copy0\kern-\wd0	
\kern-.05em\copy0\kern-\wd0
\kern-.025em\raise.0433em\box0}

\lefthead{Lazarian}
\righthead{Grain Alignment}

\begin{document}

\title{What Grain Alignment Can Tell About Circumstellar Discs and Comets}

\author{A. Lazarian}
\affil{Canadian Institute for Theoretical Astrophysics and Depatment of
Astrophysics of University of Toronto, Toronto, Canada, ON M5S 1A1\\
e-mail: lazarian@cita.utoronto.ca}
\altaffiltext{1}{Visiting Researcher, Princeton University Observatory,
Princeton, NJ 08544}

\begin{abstract}

Grain alignment theory suggests that grains should be aligned in 
circumstellar regions and the observational data available supports this
conclusion. We discuss the alignment of grains via (1) magnetic
relaxation, (2) mechanical processes, and (3) radiative torques.
We show that ferromagnetic relaxation is likely to be more important
than superparamagnetic relaxation if the dust in circumstellar
regions is similar to species recently captured in Earth atmosphere.
Outflows and stellar winds provide grain streaming along magnetic
field lines and therefore mechanical alignment competes with 
the ferromagnetic and radiative alignments. We estimate measures of
grain alignment in circumstellar regions, comets and interplanetary space
and conclude that in many circumstellar regions and in the
interplanetary space radiative torques may constitue the major
alignment mechanism which aligns grain longer axes perpendicular
to the direction of magnetic field.
Observations in submillimeter and microwave ranges are suggested as a means
of disentangling effects of  multiple scattering from those related to
aligned grains.

\end{abstract}

\keywords{ISM: Magnetic field, Polarimetry, Comets, Interplanetary Dust, 
Zodiacal Light}

\section{Introduction}

Recent years have been marked by significant advances in understanding of
grain alignment processes (see Roberge 1996, Lazarian, Goodman \& Myers 1997).
A number of new alignment mechanisms have been suggested
(e.g. Draine \& Weingartner 1996, 1997, Lazarian 1995a) and traditional
mechanisms underwent serious revision (see Lazarian \& Draine 1997). 
This process was motivated by new interstellar polarization data
(e.g. Goodman 1995, 1996) and, unfortunately, has not made the appropriate
inpact upon the areas beyond the interstellar domain.

At the same time the number of puzzling results is growing in the areas
of comet and circumstellar polarimetry, where it is customary to
believe that polarization arises from light scattering on randomly
oriented dust grains. In this paper we show that some of these puzzles
vanish if grain alignment is accounted for. 

Although models of circumstellar regions that invoke aligned grains
have been occasionally discussed in the literature (see 
Dolginov \& Mytrophanov 1976, Pudritz,
1986, 1988), their applicability was highly questionable in the
absence of the reliable grain alignment theory (see Bastien 1988). 
On the contrary,
recent theoretical advances indicate that grain alignment is likely
to be ubiquitous and therefore must be accounted for while modeling
circumstellar polarization and the polarization from comets.

In what follows we identify mechanisms of alignment that
are most efficient in circumstellar regions and in comet atmospheres
(section~2), then touch upon  the relation between grain alignment and linear and
circular polarization (section~3). In section~4
we discuss grain alignment in circumstellar regions, comets and in interplanetary space. Ways of separating the effects of multiple scattering and
those of grain alignment are discussed in section~5 and we summarize
our results in section~6. Important but more specialized
discussion of ferromagnetic versus superparamagnetic 
relaxation is given in the Appendix.

\section{Grain Alignment}

Discovered half a century ago (see Hiltner 1949, Hall 1949), grain alignment 
continues to be a tough problem for theorists. The dynamics of 
rapidly rotating dust particles is being influenced both
by numerous processes that include gaseous and ion 
bombardment, plasma effects, interactions with starlight etc.
(see more detail in Draine \& Lazarian 1998a, 1998b). Chemical processes, e.g.
H$_2$ formation that
take place on grain surfaces also influence grain dynamics (Purcell 1979,
Lazarian 1995b). Moreover, observations suggest a strong dependence of the
alignment efficiency on grain sizes. Indeed, interstellar observations
can be only explained if grains with sizes $>10^{-5}$~cm are aligned,
while smaller grains are not\footnote{This is not an exact statement
as a recent study by Lazarian \& Draine (1998) suggests that very small
grains with $a<10^{-7}$~cm may well be aligned.} (Kim \& Martin 1995).

In spite of all these difficulties substantial progress has been
recently achieved in understanding of grain alignment processes.
A list that includes six major mechanisms was presented in
Lazarian, Goodman \& Myers (1997) and a number of ``exotic'' mechanisms
have been described there as well
. Below we discuss only those of the mechanisms
that can be relevant for grain alignment in circumstellar regions and
in comet atmospheres. 
We claim that  to succeed in these environments
 the process must be fast. Therefore 
slow processes that may well work in the
interstellar medium are likely to fail in circumstellar 
regions. For instance, we do not discuss paramagnetic
alignment of suprathermal grains (Lazarian \& Draine 1997) that slowly
but steadily alignes grains over many gaseous damping times.

To characterise the alignment we use the Raylegh reduction factor
(Greenberg 1968)
\begin{equation}
R=\frac{3}{2}\langle \cos^2\beta-\frac{1}{3}\rangle~~~,
\end{equation}
where $\langle ...\rangle$ denotes the ensemble average,
$\beta$ is the angle between grain axis of maximal inertia
and the direction of alignment. We show below that often
it is magnetic field that defines direction, even for non-magnetic
alignment mechanisms.

In general, grain alignment is non-equilibrium process. Therefore
in dark clouds where ``classical'' grains are in 
thermodynamic
equilibrium with the ambient gas 
no alignment is observed (Lazarian, Goodman \& Myers 1997).
To align grains,
i.e. to decrease the enthropy of their distribution, the enthropy
of some other system (systems) must increase.

\subsection{Paramagnetic alignment}

The oldest of the alignment mechanisms is the process
of paramagnetic relaxation suggested by Davis \& Greenstein (1951) and later
modified by Purcell (1979), who observed that grains may rotate much faster
that was originally thought. To understand
the essence of this mechanism it is sufficient to consider 
a sherical grain which angular velocity makes angle $\beta$ with magnetic
field $\bf B$. The component of angular velocity perpendicular
to $\bf B$, i.e. $\omega \sin\beta$, will cause oscillating remagnetization
of the grain, while $\omega \cos\beta$ will not cause oscillations of 
magnetization. As oscillating magnetization entails dissipation, the component
$\omega \sin \beta$ decreases, while $\omega \cos\beta$ stays the same.
As the result, $\beta$ decreases.
Thus magnetic field causes anisotropy in the distribution
of grain angular momenta. As non-spherical grains  tend to rotate about 
their axes of maximal moment of inertia 
(Purcell 1979) the anisotropy in the distribution of angular
momentum is being translated into the anisotropy of the distribution of
grain longer axes.

Leaving aside the mathematical theory of alignment
(Lazarian 1997, 1998, Roberge \& Lazarian 1999), that accounts for grains
being non-spherical and internal relaxation being not complete,
we may claim that the
alignment happens on the time scale of paramagnetic relaxation, which
for ordinary paramagnetic grains is rather long, e.g.
\begin{equation}
t_{\rm al}=4\times 10^{12} K^{-1}_{-13} B^{-2}_{-5} a^2_{-5}~{\rm s}~~~,
\end{equation}
where the lower indexes used to denote the normailization values. For
instance, the $K$ function, which is the ratio of the imaginary part
of grain 
magnetic susceptibility $\chi(\omega)$ to its angular velocity $\omega$,
was normalized to $10^{-13}$~s. In other words, $K_{-13}\equiv K/(10^{-13}~{\rm s})$.
Similarly,  magnetic field is normalized by $10^{-5}$~G and grain size
$a_{-5}\equiv a/(10^{-5}~{\rm cm})$.

Grain rotation can be randomized by gaseous bombardment on time scales
\begin{equation}
t_{\rm gas}=6\times 10^{11} n_{10}^{-1} T_{\rm gas, 5000}^{-1/2} a_{-5}~{\rm s}~~~,
\end{equation}
where $n_{10}\equiv n/(3~{\rm cm}^{-3})$, $T_{\rm gas, 5000}\equiv
T_{\rm gas}/(5000~{\rm K})$. In the equation above the environmental
parameters are taken rather arbitrary
and for particular cases the more relevant values should be substituted.
Moreover the estimate for $t_{\rm gas}$ must be reduced nearly an order
of magntitude if gas is ionized (see Anderson \& Watson 1993,
Draine \& Lazarian 1998a). The latter effect is the consequence of
higher efficiency of plasma interactions with a charged grain compared
to gas-grain interactions.

To obtain efficient paramagnetic alignment $t_{\rm al}$ should be much less
than $t_{\rm gas}$.
Therefore grains with superparamagnetic and ferromagnetic inclusions 
(Jones \&
Spitzer 1967, Mathis 1986, Martin 1995, 
Draine 1996, Draine \& Lazarian 1998c) are to be considered.

How abundant ferro- and superparamagnetic 
grains in comet environment and circumstellar regions is not clear.
The presence of small $FeNi$ and $FeNiS$ inclusions in 
particles coming from the interplanetary space
has been recently reported (Bradley 1994) and this supports
the case for ``super'' grains (Goodman \& Whittet 1996). Our analysis 
of the particle image in figure~1 in Goodman \& Whittet (1996) 
indicates that most of the
inclusions are too large to exhibit superparamagnetic response for
$\omega>10^6$~s$^{-1}$\footnote{In circumstellar regions ``classical''
grains of 0.1~$\mu$m size rotate much faster due to the action of radiative
torques.}(see Appendix). However, our calculations in the Appendix prove
that the ferromagnetic response of grains with iron inclusions provides
enhancement of the paramagnetic relaxation by a factor $10^3-10^4$ if
the volume filling factor of inclusions is $\sim 0.01$ as we roughly
estimated from
the figure in Goodman \& Whittet (1996). The decrease of paramagnetic
alignment time $t_{\rm al}$
by the factor $10^4$ arising from ferromagnetic inclusions makes
$t_{\rm al}\sim 4\times 10^8$~s. This seems sufficient for circumstellar
alignment but may be slow for comet grain alignment.

\subsection{Mechanical Alignment}

Another mechanism of grain alignment stems from mechanical interaction
of grains with streaming gas. Suggested initially by Gold (1952) for
grains rotating with thermal velocities, the mechanical alignment has been 
recently proved to be efficient for grains rotating with much higher
velocities (Lazarian 1994a, 
Lazarian 1995a, Lazarian \& Efroimsky 1996, Lazarian,
Efroimsky \& Ozik 1996). Such high (suprathermal) 
velocities arise from uncompensated
quasi-regular torques, e.g. from torques arising from H$_2$ formation 
over catalytic sites on grain surface (Purcell
1975, 1979). These sites act as 
rocket engines and their action spins up the grain. The number of sites over
grain opposing surfaces, in general, is different and this causes a regular
spin-up.

The original Gold's idea is based on the observation that when a flow
of gas interacts with an elongated grain the angular momentum
 deposited with the grain tends 
to be directed perpendicular to the flow. 
Accounting for suprathermal rotation and the presence
of magnetic field makes the process of alignment a bit more involved
(see Lazarian 1995a).

The necessary condition for the mechanical alignment is the supersonic
relative motion of grains and gas. If this condition is not satisfied
isotropic gaseous bombardment randomizes grains (see eq.~(25) in
Lazarian 1997a). The rule of thumb for mechanical alignment is that the
process tends to minimize gas-grain cross section of 
interaction\footnote{This is not true, however, for the process of 
{\it alignment through friction} described in Lazarian (1995a). A
detailed discussion of the joint action of various alignment processes will
be given elsewhere.}.

It is easy to see that,
 unlike paramagnetic alignment, the mechanical one is not directly 
connected with the action of the ambient magnetic field. However, in 
many cases mechanical processes align grains either parallel or
perpendicular to the direction of magnetic field. This is the concequence
of grain rapid precession about magnetic field. Indeed, a rotating
grain aquires a magnetic moment via the Barnett effect (Dolginov \& Mytrophanov
1976, Purcell 1979) and this magnetic moment precesses in the external magnetic
field with the period
\begin{equation}
t_{\rm L}=2\times 10^{5} B^{-1}_{-5} a^{2}_{-5}~{\rm s}~~~.
\end{equation}
If $t_{\rm L}$ is much shorter than the time of mechanical alignment
$t_{\rm mech}$,
external magnetic field defines the axis of alignment.

$t_{\rm mech}$ is different for thermally and suprathermally (much
faster than thermally)
rotating grains. In the former case $t_{\rm mech}$ can be defined as
a time during which the angular momentum of a grain changes by the value
of its thermal angular momentum $J_{\rm th}=(k T_{\rm gas}/I)^{1/2}$,
where $I$ is the grain moment of inertia. In the case of suprathermally
rotating grains $t_{\rm mech}$ is the time between crossovers, i.e. moments
when grain angular velocity approaches zero and the grain flips over (see
Lazarian \& Draine 1997)\footnote{Crossovers happen due to the occasional
change of the direction of quasi-regular torques. As this direction changes
a grain first spins down then flips over and spins up.}
The time between crossovers is  approximately the sum of the 
gaseous damping time $t_{\rm gas}$ and a rather uncertain
 time of grain resurfacing (see Spitzer \& McGlynn 1979, Lazarian 1995a).
When $t_{\rm mech}\ll t_{\rm L}$ the alignment happens in respect to the
direction of gas-grain relative motion.
One could expect that in circumstellar regions both situations $t_{\rm mech}>
t_{\rm L}$ and $t_{\rm mech}< t_{\rm L}$ may be present. However, 
in many cases violent
outflows of plasma are likely to deform magnetic field lines and therefore
the correlation of the magnetic field and the direction of alignment is
expected even for $t_{\rm mech}\ll t_L$. Also note that grains carry electric charge (Martin 1972)
and therefore tend to follow magnetic field lines.

A number of processes can cause the relative grains-gas drift. Stellar
winds, outflows are examples of processes that would tend to align grains
with long axis {\it along} magnetic field lines. Ambipolar diffusion
in Roberge \& Hanany (1990) and Alfven waves in Lazarian (1994a)
were suggested as the processes that can mechanically align grains
perpendicular to magnetic field lines. In circumstellar regions and comet
atmospheres we expect mechanical alignment to happen mostly along magnetic
field lines.

\subsection{Radiative Torques}

The third mechanism that is likely to be dominant in circumstellar regions
is based on the action of radiative torques. Although mentioned first
in Dolginov (1972) and Dolginov \& Mytrophanov (1976) this
process has not been considered seriously untill very recently. Draine
\& Weigartner (1996, 1997) rediscovered the mechanism and proved
using the DDA code that radiative torques (1) can be the dominant source
of grain suprathermal rotation and that (2) these torques can align
grains
with the longer directions perpendicular to magnetic 
field. The origin of
the latter fact is not clear and this tendency contradits
to the conclusions in Dolginov \& Mytrophanov (1976)\footnote{Analytical
results in Dolginov \& Mytrophanov (1976) do not explain grain 
spin-up when the radiation is isotropically distributed. 
This fact was noted to me by Lyman Spitzer, Jr.}. Nevertheless,
treating the properties of radiative torques as established experimentally
we have to conclude that this alignment mechanism should be very
important in circumstellar regions where the radiation flux is orders
of magnitude higher than that in the interstellar environment.
Note, that even in the interstellar medium radiative torques constitue a 
major mechanism of rotation for sufficiently large, e.g. $a>10^{-5}$~cm,
grains. Within circumstellar regions with enhanced UV flux smaller
grains can be aligned radiatively. This could present a possible solution
for the recently discovered anomalies of polarization in the 2175 \Angstrom
 ~~extinction feature (see Anderson et al 1996) that has been interpreted
as the evidence of graphite grain alginment (Wolff et al 1997). If
this alignment happens in the vicinity of particular 
stars with enhanced UV flux
and having graphite grains in their circumstellar regions, this may
explain why no similar effect is observed along other lines of sight. 

Radiative torques work in unison with paramagnetic relaxation.
The situation is less clear when mechanical alignment tends to align
grains along magnetic field lines, while radiative torques act
to align grains perpendicular to magnetic field lines. 
It takes radiative torques at least a few
gaseous damping times to align grains\footnote{A peculiarity of the
radiative torque mechanism is that the gas acts as a cooling reservoir.} 
while mechanical alignment can happen in one crossover time. 
For particular angles between the direction of the incoming radiation
and magnetic field the grains perform numerous crossovers. This means
that in these situations the mechanical alignment should dominate.
The theory of crossovers in the presence of radiative torques is being
developed (Draine \& Lazarian, work in progress) and we hope to learn
soon at what conditions the mechanical alignment can win.

\section{Polarization}

Grain alignment theory can supply $R$. The observations can get Stocks
parameters. To compare observations and the theory one should related
$R$ to polarization. Because different definitions of $R$ have appeared
in the literature and confusing statements have been made in relation
to circular polarization of circumstellar origin, we find a brief discussion
of this subject appropriate.

\subsection{Linear Polarization from Aligned Grains}

For an ensemble of aligned grains the extinction perpendicular the direction
of alignment and parallel to it will be different. Therefore the 
electromagnetic wave that initially was not polarized acquires polarization.

To characterize the process quantitatively one can consider
 an electromagnetic wave propagating along the line of sight
\ofzh\ axis.
The transfer equations for the Stokes parameters
depend on the cross sections  \Cxo\ and \Cyo\ for linearly polarized
waves with the electric vector, \vecE, along the \ofxh\ and \ofyh\ directions
that are in the plane perpendicular to \ofzh\
(see Martin 1974, Lee \& Draine 1985).

To calculate  \Cxo\ and \Cyo\,
one transforms the components of \vecE\ to
a frame aligned with the principal axes of the grain and
takes the appropriately-weighted sum of the
cross sections, \Cpar\ and \Cper, for \vecE\ polarized along the grain
axes.
When the transformation is carried out and the resulting
expressions are averaged over precession angles, one finds that
the mean cross sections are
\begin{equation}
\Cxo = \Cavg + \frac{1}{3}\,R\,\left(\Cper-\Cpar\right)\,
       \left(1-3\cos^2\zeta\right)~~~,
\label{eq-2_5}
\end{equation}
where $\zeta$ is the angle between the polarization axis and the 
 \ofxh\ \ofyh\ plane,
\begin{equation}
\Cyo = \Cavg + \frac{1}{3}\,R\,\left(\Cper-\Cpar\right)~~~,
\label{eq-2_6}
\end{equation}
where $\Cavg\equiv\left(2\Cper+\Cpar\right)/3$ is the effective
cross section for randomly-oriented grains.

\subsection{Circular Polarization from Aligned Grains}

One of the ways of obtaining circular polarization is to have magnetic field
that varies along the line of sight (Martin 1972). Passing through one
cloud with aligned dust the light becomes partially linearly polirized.
On passing the second cloud with dust aligned in a different direction the
light gets circular polarized.
Literature study shows that this  effect that is well remembered 
(see Menard et al 1988), while the other process that also creates
circular polarization is frequently forgotten. We mean the process
of single scattering of light on aligned particles. Electromagnetic
wave interacting with a single grain coherently excites dipoles parallel
and perpendicular to the grain long axis. In the presence of adsorption
these dipoles get phase shift giving rize to circular polarization. 
This polarization can be observed from the ensemble of grains if
the grains are aligned. The intensity of circularly polarized
 component of radiation emerging via scattering of radiation with
$\bf k$ wavenumber on small ($a\ll \lambda$) spheroidal
particles is (Schmidt 1972)
\begin{equation}
V( {\bf e}, {\bf e}_0, {\bf e}_1)=\frac{I_0 k^4}{2 r^2}i(\alpha_{\|}
\alpha^{\ast}_{\bot}-\alpha^{\ast}_{\|}\alpha_{\bot})\left([{\bf e_0}\times
{\bf e}_1] {\bf e}\right)({\bf e}_0 {\bf e}),
\end{equation}
where ${\bf e}_0$ and ${\bf e}_1$ are the unit vectors in the directions
of incident and scattered radiation, ${\bf e}$ is the direction along
aligned axes of spheroids; $\alpha_{\bot}$ and $\alpha_{\|}$ are particle
polarizabilities along ${\bf e}$ and perpendicular to it. 

The intendity of the circular polarized radiation scattered in the
volume $\Delta \Gamma({\bf d}, {\bf r})$ at $|{\bf d}|$ from the star
and distances $|{\bf r}|$ from the observer is (Dolginov \& Mytrophanov 1978)
\begin{equation}
\Delta V ({\bf d}, {\bf r})=\frac{L_{\star} n_{\rm dust}\sigma_{V}}{6\pi |{\bf d}|^4
|{\bf r}||{\bf d}-{\bf r}|^2}R \left([{\bf d}\times {\bf r}] h\right)
({\bf d}{\bf r})\Delta \Gamma({\bf d}, {\bf r})~~~,
\end{equation}
where $L_{\star}$ is the stellar luminosity, $n_{\rm dust}$ is number 
density of dust grains and $\sigma_V$ is the cross section for
producing circular polarization, which is for small grains
is $\sigma_V=i/(2k^4)(\alpha_{\|}\alpha^{\ast}_{\bot}-\alpha^{\ast}_{\|}\alpha_{\bot})$.
According to Dolginov \& Mytrophanov (1978) circular polarization arising 
from single scattering on aligned grains
can be as high as several percent for metallic or graphite particles,
which is much more than one expects
from the process of varying magnetic field direction along the line of
sight.

\section{Particular cases}

\subsection{Circumstellar Regions}

Multiple scattering has been used to explain polarization
arising from circumstellar regions (see Bastien 1988, 1996). At the same
time it is obvious that in the presence of radiation and magnetic
field, grains in circumstellar envelops must be aligned perpendicular
to magnetic field. For the stars that exhibit outflows and intensive
stellar winds, numerical models (see Netzer \& Elitzur 1994) predict
a supersonic relative drift of grain and gas and this should result
in mechanical alignment. In circumstellar environments
the grain rotation temperature
is likely to be much higher than its body temperature. Therefore
results for mechanical obtained in 
Lazarian (1994a) and Lazarian (1995a) are applicable.
This entails $R\sim -0.3$ for both prolate and oblate grains with
grain long axis along the outflow direction. The uncertainty involved,
as we have mentioned earlier, is related to the absence of the theory of
radiative crossovers. We may claim that our estimate of $R$ is valid
for sufficiently small (e.g. $a< 5\times 10^{-6}$~cm) grains, while
for larger grains the situation is unclear as yet.

If grains have superparamagnetic or ferromagnetic inclusions and for
radiative torques the alignment tends to be nearly perfect (i.e. $R\sim 1$)
with the logner grain dimensions perpendicular to magnetic field lines.
If, however, a grain with ferromagnetic inclusions (e.g. 
``Goodman-Whittet grain'' discussed above) is subjected to
streaming along field lines, it will be aligned perpendicular
to magnetic field lines as the magnetic relaxation time is typically
shorter than that for mechanical alignment. We tend to believe
that grain alignment with grain longer axes perpendicular to 
magnetic field and $R\sim 1$ can be a rule for circumstellar regions.
Future research should test this conjecture.

The examples above indicate that future modeling of circumstellar regions
should include aligned grains. Whether multiple scattering or 
dichroic adsorption is dominant should be decided
by quantitative comparison of the simulations that include both 
effects and observations. Submillimeter polarimetry will be helpful
for establishing grain alignment in circumstellar regions (see below).

\subsection{Comets}

Polarization from comets has been long known to exhibit anomalies
(see Martel 1960) that motivated a conjecture that grains may be aligned
in the comet atmospheres (see Dolginov \& Mytrophanov 1976). Later
studies of linear and circular polarization from Halley and Hale-Bopp
comet (Beskrovnaja et al 1987, Ganesh et al 1998) seem to support this
conclusion.

The alignment mechanism operating in comet heads should be really fast. 
Indeed, dust particles spend only $\sim 10^5$~s crossing a comet head.
Unless magnetic field in the comet head is extremely high (e.g. $> 10^{-2}$~G)
the ferromagnetic relaxation fails to provide the alignment. In comet
heads grains are likely to disaggregate and change their shape rather
rapidly. This should mitigate the importance of raditative torques
that will change their direction with the change of grain shape. At the
same time, dramatic changes of grain shapes on the timescale $t_{\rm mech}$
wash out the distinction
between prolate and oblate grains and hinder the mechanical alignment as
well.

We believe that outflowing gases can be important for grain alignment at
the comet head.
Calculations in Probstein (1969) indicate that the relative velocities
of dust and gas are supersonic. We expect the alignment for
thermally rotating grains to be small (e.g. $R\sim -0.1$) and to happen
in respect to the outflow direction. Higher degrees of alignment
are possible (e.g. $R\sim -0.3$) if grains rotate suprathermally.
Indeed, both radiative torques and assymetry in the gas evaporation from grain
surface may contribute to suprathermal rotation. Very large dust particles
(e.g. $a>10^4$~cm) may be aligned by a weathercock mechanism discussed
in Lazarian (1994b).

Later, in the outer parts of comet
coma and in its tail the alignment via radiative torques and interaction
with solar wind should be important. $R$ approaching unity is
attainable in the former case.
Quantitative modeling of the
grain alignment in comets is under way (Bastien \& Lazarian, work in
progress).

\subsection{Zodiacal Light}

Zodiacal Light, i.e. solar light reflected from the interplanetary dust
particles, is partially polarized. Greenberg (1970) suggested that
 interplanetary grains could be aligned. Later on similar ideas were
discussed by e.g. 
Wolstencroft \& Kemp (1972) and Dolginov \& Mytrophanov (1978).

Greenberg (1970) worried that interplanetary particles can be sputtered
quicker than be aligned by solar wind. However, his arguments ignore
important plasma interactions and ion focusing effect (see Draine \& Lazarian
1998b) that make transfer of angular momentum from solar wind to grains much
more efficient. Thus mechanical alignment is concivable ($R\sim -0.3$)
with grain long axis along magnetic field lines.

The alignment by radiative torques and via ferromagnetic relaxation
are possible as well. If large silicate grains that produce most of the
linear polarization are aligned along magnetic field lines, while a
possible population of absorbtive iron grains that would account for
most of the circular polarization are aligned perpendicular to interplanetary
magnetic field, quite complex picture of polarization may arise. 
However, it is likely that mechanical alignment is most important for
small ($a<5\times 10^{-6}$~cm) grains, while larger grains are being aligned
by radiative torques. Then both small iron grains and large silicate
ones are being aligned with long axes perpendicular to the direction of
the interplanetary magnetic field. Potentially,
studies of Zodiacal Light can bring a lot of information about magnetic
field structure and its variability in the Solar neighborhood. 

The interplanetary magnetic field, as well as those of circumstellar regions
and comets, is not stationary. In fact it undergoes variations on a
whole range of time scales. If the variations are long compared to the
Larmor period $t_L$ they are adiabatic in the sence that the angle
between grain angular momentum and $\bf B$ is preserved. Therefore 
time variations of the Zodiacal Light can provide important
information on the magnetic
variability up to the scale $t_L$.

\section{Future Work}

It is often  difficult to separate the effects of multiple stattering from
the effects of grain alignment. One of the alluring possibilities is
to observe at longer wavelengths, where the effects of multiple stattering
are negligible. Polarimetry at submillimeter and longer wavelengths should
help constructing adequate models of polarized light transfer
in circumstellar regions and comets and unrevel magnetic field
structure in these regions. 

Our discussion above was centered on the issue what ``classical''  or
sufficiently large grains can tell us. It looks, however, that very
small grains can make a valuable input as well. Recent experiments to map
cosmic microwave background, e.g. Kogut et al (1996), Oliveira-Costa
et al (1997) and Leitch et al (1997), have revealed a new component of
galactic microwave emission at 14 - 90 GHz. This component was identified by 
Draine \& Lazarian (1998a) with the dipole emission from small 
($a<10^{-7}$~cm) rotating grains. Lazarian \& Draine (1998) predicted
that such grains can be aligned and that this should result in anomalous
emission being partially polarized. This opens a new valuable window
for interstellar and circumstellar studies. An important feature of
the relaxation mechanism suggested is that it stays efficient even
when ``classical'' grains are in thermodynamic equilibrium with the
ambient gas and are randomly oriented. Thus the progress in grain alignment
theory presents new tools for observers.

\section{Conclusions}

The principal results of this paper are as follows:

The application of the results obtained in grain alignment theory to comets
and circumstellar regions suggest that the dust should be aligned there.
Three most important alignment mechanisms are (1) radiative torques,
(2) mechanical alignment, (3) ferromagnetic and superparamagnetic
relaxation. Observational data supports the conjecture that the dust is
aligned in circumstellar regions and comets. Therefore numerical codes that 
describe radiation transfer in young
stellar objects and evolved stars should be modified to account for
dust alignment.

The analysis of the images of the dust
particles coming from the interplanetary space testify that the ferromagnetic
relaxation, rather that superparamagnetic relaxation is likely. The calculated
enhancement of the relaxation (compared to that in paramagnetic grains) 
is $\sim 10^4$ and is sufficiently large to enable the efficient
alignment of circumstellar dust with ferromagnetic inclusions.

Mechanical alignment and radiative torques compete in aligning grains,
(along and perpendicular magnetic field lines, respectively) in the regions 
of outflows. When streaming velocities are supersonic small grains 
($a<5\times 10^{-6}$~cm) without 
ferromagnetic inclusions are to be aligned with long axes
parallel to magnetic field
lines, while those with ferromagnetic inclusions are to be aligned
with long axes perpendicular to the field lines. The situation
is still unclear with large ($a>10^{-5}$~cm) grains, but we conjecture
that at least in circumstellar regiona and interplanetary space grains
are aligned with long axes perpendicular to magnetic field.

Both linear and circular polarization provide a valuable input on 
magnetic fields in circumstellar regions, comet atmospheres and in the Solar
neighborhood. Measurements at submillimeter wavelenghs can disentangle
effects of multiple scattering from those of grain alignment. In particular
cases when large grains are not aligned it is advisable
to use microwave polarimetry that is sensitive to the alignment of
tiny ($a<10^{-7}$~cm) grains.

{\bf Acknowledgements}
\acknowledgements
I am grateful to Pierre Bastien, Bruce Draine, Alyssa Goodman and 
Peter Martin for helpful discussions and happy to acknowledge the 
support of NASA grant NAG5 2858 and CITA Senior Research Fellowship.

\appendix
\section{Ferromagnetic and Superparamagnetic Susceptibilities}

How superior can be ``supergrains'' in terms of paramagnetic relaxation?
To answer this question we consider   
iron inclusions. It is well known that 
small iron particles particles are superparamagnetic (Morrish 1980).
If iron forms clusters containing  $N$ atoms the zero-frequency magnetic susceptibility of a grain 
increases $N$ times compared with a grain where the same amount of iron is 
uniformely distributed within a diamagnetic lattice (Draine 1996):
\begin{equation}
\chi(0)_{\rm super}\approx N \chi_{\rm param}~~~,
\end{equation}
where
\begin{equation}
\chi_{\rm param}\approx 0.04 f_{\rm p} \left(\frac{n_{\rm tot}}{10^{23}~{\rm cm}^{-3}}\right)\left(\frac{p}{5.5}\right)^2 \left(\frac{15~{\rm K}}{T_{\rm grain}}\right)~~~.
\end{equation}
Above $p\mu_B$ is the magnetic moment of a paramagnetic ion, 
$\mu_B=e\hbar/2m_ec$ is
the Bohr marneton and $n_p=f_{\rm p} n_{\rm tot}$ is the number of paramagnetic
ions in a grain with density $n_{\rm tot}$.
Within interstellar grains $f_{\rm p}$ is of the order of $0.1$. We 
use this value as a rough  estimate for the circumstellar and cometary dust.

How large can be a particle to exhibit superparamagnetic response in
oscilating magnetic field with frequency $\omega$ depends on the thermally
activated relaxation rate
\begin{equation}
\tau_{\rm activ}\approx \nu_0 \exp[-N T_{\rm activ}/T_{\rm grain}]
\end{equation}
where $T_{\rm activ}\approx 0.011$~K and $\nu_0\approx 10^9$~s for
Fe particles (Bean \& Livingston 1959).

For $\tau_{\rm activ}\omega\ll 1$ $K(\omega)_{\rm super}$ that is equal to the
imaginary part of $\chi(\omega)_{\rm super}/\omega$ is approximately
$\chi_{\rm super} \tau_{\rm activ}$ (Spitzer 1978). 
When $\tau_{\rm activ}\omega>1$
$K(\omega)_{\rm super}$ rapidly decreases with $\omega$ 
(Jones \& Spitzer 1968, Draine \& Lazarian 1998c). 
It is easy to show
that the number of iron atoms should not exceed $3\times 10^3$ to enable
efficient paramagnetic 
relaxation of grains rotating faster than $10^5$~s$^{-1}$. Therefore the 
maximal value of $K_{\rm super}$ is  approximately $3\times 10^3  
\chi_{\rm param}\tau_{\rm activ}\approx 3\times 10^{-6}\chi_{\rm param}$.
This value should be compared to
$K_{\rm param}$ which is approximately $ 3\times 10^{-11}
\chi_{\rm param}$ (see Draine 1996). All in all, the maximal
increase of relaxation due to superparamagnetism is given by
a factor $10^5$. 

The latter factor of the relaxation enhancement is frequently
quoted in the literature without mentioning that, first of all, this
is an upper limit for superparamagnetic relaxation enhancement
and, even more important, that inclusions of larger
size do not exhibit superparamagnetic response for $\omega > 10^5$~s$^{-1}$.
The minimal number of paramagnetic atoms that make up a superparamagnetic
inclusion is uncertain. We know that inclusions with more than 20 atoms 
do  exhibit superparamagnetism (Billas,
Chatelain \& de Heer 1994). For 30 atom inclusions the superparamagnetic
relaxation is $10^3$ times enhanced. Inclusions
with more than $3000$ atoms will exhibit ferromagnetic 
properties.

The magnetic susceptibility of large particles
follows from the solution of the Bloch equations (see Pake 1962)
\begin{equation}
\chi(\omega)\approx \chi_{Fe} (0)\frac{\omega}{1-(\omega/\omega_0)^2-i\omega \tau}~~~,
\label{chi}
\end{equation}
where $\chi(0)$ is the zero frequency magnetic susceptibility and
$\omega_0$ and $\tau$ are two parameters that have the meaning of the
characteristic frequency and time. To approximate experimental results
available (see Epstein 1954) one can assume that $\chi_{Fe}\approx 10$,
 $\tau\approx 10^{-9}$~s and $\omega_0\approx
10^{10}$~s$^{-1}$ (Draine \& Lazarian 1998c). The susceptibility of
a grain with large inclusions may be estimated using effective medium
theory (Bohren \& Huffman 1984). For a small volume filling factor $\phi\ll1$
\begin{equation}
\chi_{eff}\approx \frac{\phi \chi(\omega)}{1+2\pi \chi(\omega)}~~~,
\end{equation}
where $\chi(\omega)$ is given by Eq.~(\ref{chi}). Therefore for the 
volume filling factor of $0.01$ the efficiency of relaxation for
grains with ferromagnetic inclusions is approximately $10^4$ times that
of a paramagnetic grain. More elaborate calculations show that 
grains with single domain inclusions 
exhibit susceptibilities which are a factor 5 smaller than those
found above for the multidomain $Fe$ inclusions (Draine \& Lazarian 1998c).

\end{document}